\documentclass[prd,a4paper, preprintnumbers, amsmath, multicol]{revtex4}
\usepackage{latexsym}
\usepackage{epsfig}
\usepackage{amsmath}
\usepackage{amssymb}
\usepackage{graphicx}
\usepackage{bm}% bold math
\usepackage{color}
\definecolor{rosso}{cmyk}{0,1,1,0.4}
\definecolor{rossos}{cmyk}{0,1,1,0.55}
\definecolor{rossoc}{cmyk}{0,1,1,0.2}
\definecolor{blu}{cmyk}{1,1,0,0.3}
\definecolor{blus}{cmyk}{1,1,0,0.6}
\definecolor{bluc}{cmyk}{1,1,0,0.1}
\definecolor{verde}{cmyk}{0.92,0,0.59,0.25}
\definecolor{verdec}{cmyk}{0.92,0,0.59,0.15}
\definecolor{verdes}{cmyk}{0.92,0,0.59,0.4}

\makeatletter

\newcommand{\beq}{\begin{equation}}
\newcommand{\eeq}{\end{equation}}
\newcommand{\bea}{\begin{eqnarray}}
\newcommand{\eea}{\end{eqnarray}}
\newcommand{\ba}{\begin{array}}
\newcommand{\ea}{\end{array}}
\newcommand{\bi}{\begin{itemize}}
\newcommand{\ei}{\end{itemize}}
\newcommand{\bn}{\begin{enumerate}}
\newcommand{\en}{\end{enumerate}}
\newcommand{\bc}{\begin{center}}
\newcommand{\ec}{\end{center}}

\newcommand{\gsim}{\lower.7ex\hbox{$\;\stackrel{\textstyle>}{\sim}\;$}}
\newcommand{\lsim}{\lower.7ex\hbox{$\;\stackrel{\textstyle<}{\sim}\;$}}

\begin{document}

\author{D. Bettoni$^{\color{blue}{1}}$   , 
P. Dalpiaz$^{\color{blue}{1}}$ ,
 P. F. Dalpiaz$^{\color{blue}{1}}$,
 M. Fiorini$^{\color{blue}{1}}$, 
I. Masina$^{\color{blue}{1}}$, 
G. Stancari$^{\color{blue}{1,2}}$}

\affiliation{
$^{\color{blue}{1}}$ 
{Dip.~di Fisica dell'Universit\`a di Ferrara and INFN Sez.~di Ferrara, Via Saragat 1, I-44122 Ferrara, Italy}\\ 
$^{\color{blue}{2}}$ {Fermi National Accelerator Laboratory, Batavia,IL 60510, USA }\\}

\title{\Large \color{red} \bf Experimental proposal to study the excess at $M_{jj}=150$ 
GeV presented by CDF at Fermilab}

\baselineskip=15pt

\setcounter{page}{1}

\vskip 1cm

\begin{abstract}
\begin{center} {\bf Abstract} \end{center}  

We propose an experimental test to verify the unexpected excess at
$Mjj=150$ GeV presented by the CDF
collaboration in the invariant mass distribution of jet pairs produced in
association with a W boson.
 We propose a formation experiment in which the energy range of the 
$M_{jj}$ excess is scanned with proton-antiproton interactions at the Tevatron.
\end{abstract}

\maketitle
%\tableofcontents

%%%%%%%%%%%%%%%%%%%%%%%%%%%%%%%%%%%%%%%%%%%%%%%%%%%%%%%%%%%%%%%%%%%

\section*{}

The excess at $M_{jj}=150$ GeV in the invariant mass distribution of jet pairs produced in
association with a W boson,  
 recently presented by CDF collaboration \cite{Aaltonen:2011mk} was a big surprise. 
The reported  cross section  is $300$ times larger than expected for this channel in the frame 
of the Standard Model.  This result needs an experimental confirmation. 
The CDF collaboration has already data to double the statistics and when the analysis is completed we will 
know whether the effect is confirmed.

We propose an independent experimental validation of this effect,
scanning the observed bump in $M_{jj}$, {\it in formation} in $p \bar p$ interactions,
setting the Tevatron energy in the range of the $M_{jj}$ excess.

An important feature of formation experiments is that the energy resolution is given by the momentum resolution 
of the beams. The spectrometer is used only in the trigger, to select the events.
This experimental technique  has been successfully exploited by many experiments 
for the study of hadron spectroscopy: in particular it was used to study  the charmonium spectrum  
by three experiments: R704 \cite{Baglin:1986br} at the CERN ISR, and E760 and E835 \cite{Garzoglio:2004kw}
at the Fermilab antiproton accumulator.
The same technique will be used by the PANDA experiment at the future  FAIR facility \cite{Lutz:2009ff}. 
  
This kind of experiments is analogous and complementary to the study of resonances {\it in formation } in $e^+ e^-$
colliders.  It must be noted, however, that in $e^+e^-$ annihilation direct formation is limited to the 
states with the quantum numbers of the photon
($J^{PC}=1^{--}$), whereas in $p\bar p$ formation all the states with 
any (non-exotic) quantum numbers can be formed directly. As a consequence,  in $p\bar p$ annihilation 
the background is higher with respect to 
$e^+e^-$. Furtheremore, the cross section, that is $1\mu barn$ in  
the reaction $p\bar p \rightarrow J/\Psi$,  decreases 
very rapidly with the mass of the resonances if the photon exchange dominates the process.
 
On the other hand, the hadronic 
background problem can be very effectively overcome by looking at specific final states.
In fact, this technique is very useful when the widths of the states 
analyzed are narrow and the signature clear,  allowing  a 
very precise measurement of the masses and widths.
 \newpage

\begin{figure}[]
\begin{center}%\begin{tabular}{c}
\includegraphics[angle=90,width=15cm]{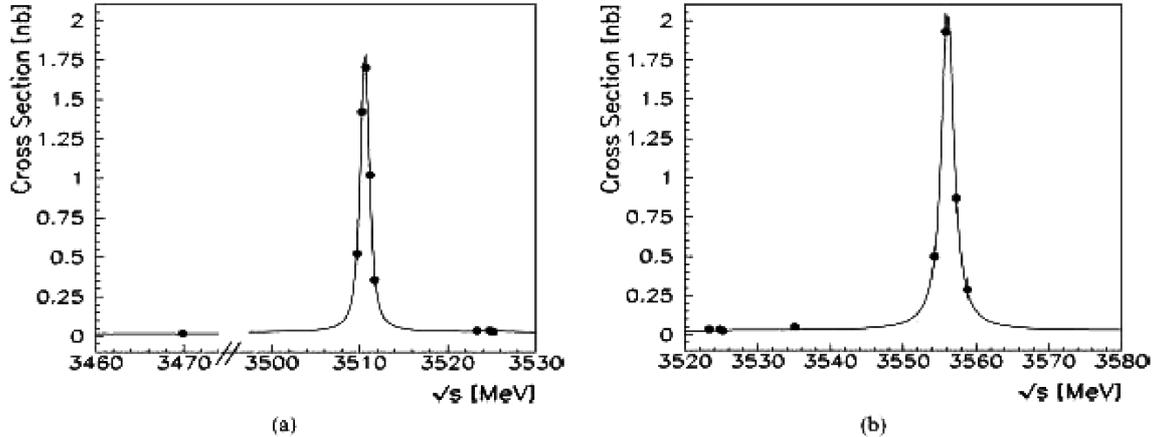} %\end{tabular}
\end{center}\vspace*{-0.5cm} 
\caption{ E835 measurements [5] of the cross section as a function of the
center-of-mass energy for the $\chi_{c1}$  (a) and  $\chi_{c2}$ (b) charmonium states
via the process   $p\bar p \rightarrow \chi_c \rightarrow J/\Psi + \gamma \rightarrow e^+ + e^- +\gamma $.}
\label{fig-cc}
\end{figure}

This is clearly demonstrated  in Fig. \ref{fig-cc}, 
which shows the E835  scans of
the   $\chi_{c1}$  and  $\chi_{c2}$   charmonium states via the process 
   $p\bar p \rightarrow \chi_c \rightarrow
J/\Psi + \gamma \rightarrow e^+ + e^- +\gamma$, where the reaction is triggered only 
by the electron pair \cite{Andreotti:2005ts}. The
experiment was performed at the Fermilab antiproton accumulator with
an internal hydrogen jet target. The energy scans were carried out
varying the antiproton beam momentum in small steps. The excellent
beam momentum resolution of $2 \times 10^{-4}$ allowed E835 to produce the most
precise measurements of the masses and widths of these very narrow states.
The figure also shows the excellent signal-to-noise ratio which can be
obtained in this kind of formation experiments triggering on exclusive
final states with a clear signature.

%%%%%%%%%%%%%%%%%%%%%%%%%%%%%%%%%%%%%%%%%%%%%%%%%%%%%%%%%%%%%%%%%%%%%%%%%
The suggestion to scan  the observed bump in $M_{jj}$, 
setting the Tevatron energy in the range of the $M_{jj}$ excess, is therefore straightforward.

We observe, however, that the Tevatron energy ranges between 300GeV and 2TeV. 
At the Tevatron injection energy of $150$GeV per beam, one can expect an initial instantaneous 
luminosity of $3\times 10^{31}$cm$^{-2}$s$^{-1}$ and an average luminosity of $6$ events/(pb week), 
taking into account beam lifetimes, emittance growth, and turn-around time. 
At injection, the relative beam momentum spread is $5 \times 10^{-4}$.  This spread
means 300MeV energy resolution. 

At lower  energies,  as 150 GeV , it is possible to operate the Tevatron, but a particular machine development
is necessary. Also the forseen luminosity decreases by a factor 2.

Many theoretical interpretations of the $M_{jj}$ excess have appeared in the last month.  
One of the first, proposed by Eichten et al. \cite{Eichten:2011sh}, consider the $M_{jj}$ excess 
in the Technicolor framework. The authors identify the $M_{jj}$ excess as the techni-pion, $\pi_T$. 
Within this model the $\pi_T$ is the decay product of a resonance with $I^G J^PC = 1^+ 1^{--}$,  
the techni-rho vector meson, $\rho_T \rightarrow W \pi_T$, with $M_{\rho_T}= 290$ GeV. 
The $\pi_T$ decays in two jets. This model takes into account the fact that the $M_{jj}$ excess is not seen 
in experiments with leptons, the quark b are absent in the decay jets and the cross section is $300$ times 
larger than the expectation of the Standard Model.

We propose to test this theoretical hypothesis, searching for the formation reaction
of the           techni-rho vector meson:
 $  p\bar p \rightarrow \rho_T  $,    $\rho_T \rightarrow W \pi_T$ . This is possible,
setting the Tevatron near the injection energy 
and scanning around $290$GeV, triggering on $[e(\mu)]$ produced in the $W$ decay, and the 
jet pairs $[jj]$ from
the techni-pion, $\pi_T$, decay. 
Notice that the $W$ produced in the decay is nearly 
monochromatic and longitudinally polarized. 
The $e$ and $\mu$ produced in the $W$ semileptonic decays have then an  energy spectrum, approximately 
proportional to $x (1-x)$, where $x=E/E_W$, $E$ is the energy of the $e(\mu)$ and $E_W$ is the energy
of the $W$. These features, a clear signature of the searched events,
 used in the trigger and in the susequent analysis, will greatly improve the background reduction.  

Obviously, the cross section of the hypothethical reaction   $  p\bar p \rightarrow \rho_T  $ is unknown.
From CDF evidence,  however,  we know that at $2 TeV$ energy, the cross-section  should be around  $4 pb$.
Taking this cross-section value for our process, CDF or D0 will collect  each 24  events a week.
 If the techni-rho is, as expected,  narrower than $1$GeV, it is possible to overcome the background
also with some lower cross section.

%%%%%%%%%%%%%%%%%%%%%%%%%%%%%%%%%%%%%%%%%%%%%%%%%%%%%%%%%%%%55

The experimental test could be done at Tevatron using CDF and D0 detectors without modifications, 
with a reliable luminosity monitor.   It may be observed that, in this case, the consumption of 
electric power and helium is  lower than 
in the normal running at high Tevatron energy.

When it will be possible to set the Tevatron energy at $150$GeV,
and scan around  
the CDF observed $M_{jj}$ bump,   triggering on collinear jets $[jj]$,
a direct indipendent validation of the CDF  observed bump in $M_{jj}$ could be
realized.

 Concluding, we observe that the Tevatron has a unique possibility to test this
 new physics in short times.

%\newpage

%%%%%%%%%%%%%%%%%%%%%%%%%%%%%%%%%%%%%%%%%%%%%%%%%%%%%%%%%%%%%%%%%%%%%%%%%%%%%%%%%%%%%%%%%%%%%%%%%%%%%%%%%%

\end{document}